\documentclass[aps,prl,twocolumn,superscriptaddress,showpacs,showkeys,amssymb,amsmath]{revtex4-1}
\usepackage{graphicx}

\newcommand{\V}[1]{\mathbf{#1}} 
 
\newcommand\Alfven{Alfv\'en }
\newcommand\Alfvenic{Alfv\'enic }
\newcommand{\figref}[1]{Fig.~\ref{#1}}

\newcommand{\xhat}{\mbox{$\hat{\mathbf{x}}$}}

\usepackage{color}

\begin{document}

\title{Toward Astrophysical Turbulence in the Laboratory}

\author{G.~G. Howes}
\affiliation{Department of Physics and Astronomy, University of Iowa, Iowa City, 
Iowa 52242, USA.}
\author{D.~J.~Drake}
\affiliation{Department of Physics and Astronomy, University of Iowa, Iowa City, 
Iowa 52242, USA.}
\affiliation{Department of Physics, Astronomy, and Geosciences, Valdosta State University, Valdosta, Georgia 31698, USA.}
\author{K.~D.~Nielson}
\affiliation{Department of Physics and Astronomy, University of Iowa, Iowa City, 
Iowa 52242, USA.}
\author{T.~A.~Carter}
\affiliation{Department of Physics and Astronomy, University of California, Los Angeles, California 90095-1547, USA.}
\author{C.~A.~Kletzing}
\affiliation{Department of Physics and Astronomy, University of Iowa, Iowa City, 
Iowa 52242, USA.}
\author{F.~Skiff}
\affiliation{Department of Physics and Astronomy, University of Iowa, Iowa City, 
Iowa 52242, USA.}

\date{\today}

\begin{abstract}
Turbulence is a ubiquitous phenomenon in space and astrophysical
plasmas, driving a cascade of energy from large to small scales and
strongly influencing the plasma heating resulting from the dissipation
of the turbulence.  Modern theories of plasma turbulence are based on
the fundamental concept that the turbulent cascade of energy is caused
by the nonlinear interaction between counterpropagating \Alfven waves,
yet this interaction has never been observationally or experimentally
verified. We present here the first experimental measurement in a
laboratory plasma of the nonlinear interaction between
counterpropagating \Alfven waves, the fundamental building block of
astrophysical plasma turbulence.  This measurement establishes a firm
basis for the application of theoretical ideas developed in idealized
models to turbulence in realistic space and astrophysical plasma
systems.
\end{abstract}

\maketitle 

\emph{Introduction.}---Turbulence profoundly affects many space and 
astrophysical plasma environments, playing a crucial role in the
heating of the solar corona and acceleration of the solar wind
\cite{McIntosh:2011},  the dynamics of the interstellar medium
\cite{Armstrong:1981,Armstrong:1995,Gaensler:2011}, the regulation of
star formation \cite{Rickett:1990}, the transport of heat in galaxy
clusters \cite{Peterson:2006}, and the transport of mass and energy
into the Earth's magnetosphere
\cite{Sundkvist:2005}. At the large length scales and low frequencies
characteristic of the turbulence in these systems, the turbulent
motions are governed by the physics of \Alfven waves
\cite{Alfven:1942}, traveling disturbances of the plasma and magnetic 
field.  Theories of Alfv\'enic turbulence based on idealized models,
such as incompressible magnetohydrodynamics (MHD), suggest that the
turbulent cascade of energy from large to small scales is driven by
the nonlinear interaction between counterpropagating \Alfven waves
\cite{Kraichnan:1965,Sridhar:1994,Goldreich:1995,Boldyrev:2006}. 
However, the applicability of this key concept in the moderately to
weakly collisional conditions relevant to astrophysical plasmas has
not previously been observationally or experimentally demonstrated.
Verification is important because the distinction between the two
leading theories for strong MHD turbulence
\cite{Goldreich:1995,Boldyrev:2006} arises from the detailed nature
of this nonlinear interaction. Furthermore, verification is required
to establish the applicability of turbulence theories, utilizing
simplified fluid models such as incompressible MHD, to the weakly
collisional conditions of diffuse astrophysical plasmas.

Several reasons make it unlikely that the nonlinear interaction
between counterpropagating \Alfven waves can ever be verified using
observations of turbulence in astrophysical environments: the spatial
resolution achievable in astrophysical observations is insufficient,
\emph{in situ} spacecraft measurements yield information at only a
single or a few spatial points, and the broad spectrum of turbulent
modes confounds attempts to identify the transfer of energy from two
nonlinearly interacting
\Alfven waves to a third wave. Only experimental measurements in the
laboratory can achieve the controlled conditions and high spatial
resolution necessary. The unique capabilities of the Large Plasma
Device (LAPD) at UCLA \citep{Gekelman:1991}, designed to study
fundamental plasma physics processes, make possible the first
laboratory measurement of the nonlinear wave-wave interaction
underlying \Alfvenic turbulence.

This Letter presents the first laboratory measurement of the nonlinear
interaction between counterpropagating \Alfven waves, the fundamental
building block of astrophysical plasma turbulence.  The properties of
the nonlinear daugther \Alfven wave are predicted from incompressible
MHD theory. The experimental setup and procedure are
outlined. Analysis of the experimental results demonstrate a
successful measurement of the nonlinear interaction between
counterpropagating
\Alfven waves.

\emph{Theory.}---Modern theories of anisotropic \Alfvenic plasma turbulence are based
on several key concepts derived from the equations of incompressible
MHD.  These equations can be expressed in the symmetric form
\cite{Elsasser:1950},
\begin{equation} 
\partial \V{z}^{\pm}/\partial t
 \mp (\V{v}_A \cdot \nabla )\V{z}^{\pm} 
=- (\V{z}^{\mp}\cdot \nabla)  \V{z}^{\pm} - \nabla p/\rho_0,
\label{eq:elsasser}
\end{equation}
where the magnetic field is decomposed into $\V{B}=\V{B}_0+ \delta
\V{B} $, $\V{v}_A =\V{B}_0/\sqrt{\mu_0\rho_0}$ is the \Alfven velocity 
due to the equilibrium field $\V{B}_0$, $p$ is total pressure (thermal
plus magnetic), $\rho_0$ is mass density, and $\V{z}^\pm = \V{v}_\perp
\pm \delta \V{B}_\perp/\sqrt{\mu_0\rho_0}$ are the Els\"asser 
fields of the \Alfven waves which are incompressible, so that $\nabla
\cdot \V{z}^{\pm} =0$.  The Els\"asser field $\V{z}^{+}$ ($\V{z}^{-}$)
represents an \Alfven wave traveling down (up) the mean magnetic
field.  The second term on the left-hand side of \eqref{eq:elsasser}
is the linear term representing the propagation of the Els\"asser
fields along the mean magnetic field at the \Alfven speed, the first
term on the right-hand side is the nonlinear term representing the
interaction between counterpropagating waves, and the second term on
the right-hand side ensures incompressibility \cite{Goldreich:1995}.

Consider the nonlinear interaction between two plane \Alfven waves
with wavevectors $\V{k}_1$ and $\V{k}_2$, each with non-zero
components both parallel and perpendicular to the equilibrium magnetic
field. The mathematical form of the nonlinear term
in \eqref{eq:elsasser} requires two conditions for nonlinear
interaction to occur: (a) both $\V{z}^{+} \ne 0$ and $\V{z}^{-} \ne
0$, so the two waves must propagate in opposite directions along the
magnetic field, implying $k_{\parallel 1}$ and $k_{\parallel 2}$
have opposite signs \cite{Iroshnikov:1963,Kraichnan:1965}; and, (b)
the polarizations of the perpendicular wave magnetic fields are not
parallel, implying $\V{k}_{\perp 1} \times \V{k}_{\perp 2} \ne
0$. These properties dictate that the fundamental building block of
plasma turbulence is the nonlinear interaction between perpendicularly
polarized, counterpropagating \Alfven waves.

For sufficiently small amplitudes, the terms on the right-hand side of
\eqref{eq:elsasser} are small compared to the linear propagation term, producing a
state of \emph{weak MHD turbulence} \cite{Sridhar:1994}. Note that the
linear term in \eqref{eq:elsasser} has no counterpart in
incompressible hydrodynamics, eliminating the possibility of weak
turbulence, a fundamental distinction between incompressible
hydrodynamic and incompressible MHD systems.  In the weak MHD
turbulence paradigm, two counterpropagating \Alfven waves interact
nonlinearly to transfer energy to a third wave.  This is the
fundamental interaction underlying the cascade of energy to small
scales in plasma turbulence. Solving for the nonlinear evolution using
perturbation theory demonstrates that the nonlinear three-wave
interaction, averaged over many wave periods, must satisfy the
constraints
\begin{equation}
\V{k}_1+ \V{k}_2 = \V{k}_3 \quad \mbox{ and } \quad \omega_1 + \omega_2 = \omega_3,
\label{eq:constraints}
\end{equation}
equivalent to the conservation of momentum and energy
\citep{Sridhar:1994,Ng:1996,Galtier:2000}. Given  the dispersion 
relation for \Alfven waves, $\omega=|k_\parallel| v_A$, and the
requirement for counterpropagating waves, the only nontrivial solution
to these equations has either $k_{\parallel 1}=0$ or $k_{\parallel
2}=0$. \citep{Shebalin:1983}

The frequency constraint in \eqref{eq:constraints} ceases to hold when
the interaction spans only a fraction of a wave period.  For such
brief interactions, energy is transferred nonlinearly to a daughter
mode at the instantaneous rate given by the nonlinear term in
\eqref{eq:elsasser}.  In the laboratory, this can be accomplished by
interacting a high-frequency \Alfven wave $\V{k}_{1}$ with a
counterpropagating \Alfven wave $\V{k}_{2}$ of much lower frequency,
such that its parallel wavelength is much longer than the length over
which the waves interact.  The physical effect of the low-frequency
wave is to generate a shear in the equilibrium magnetic field,
producing an effective $k_{\parallel 2}=0$ component to the
interaction.  The predicted nonlinear product is an \Alfven wave
$\V{k}_{3}$ with the properties $k_{\parallel 3} = k_{\parallel 1}$
and $\V{k}_{\perp 3} =\V{k}_{\perp 1}+ \V{k}_{\perp 2}$.

\emph{Experiment.}---The first experiment to verify this interaction 
in the laboratory was performed on the LAPD \citep{Gekelman:1991}
using a background axial magnetic field of 800~G to confine a hydrogen
plasma in a cylindrical column of 16.5~m length and 40~cm
diameter. The plasma discharge exists for approximately 11~ms with a
repetition rate of 1~Hz.  The electron temperature, $T_e=5.0$~eV, and
density, $n=10^{12}$~cm$^{-3}$, were determined using a swept Langmuir
probe, with a microwave interferometer used to calibrate density.  The
ion temperature, $T_i=1.25$~eV, was estimated from previous
interferometric measurements. For these parameters, the electron-ion
collision frequency \citep{Braginskii:1965} is 3~kHz and the
characteristic perpendicular scale is given by the ion sound Larmor
radius $\rho_s =\sqrt{T_e/m_i}/\Omega_i = 0.29$~cm, where $\Omega_i$
is the ion cyclotron frequency.

\begin{figure}
\resizebox{86mm}{!}{\includegraphics*[1.15in,2.8in][8.02in,6.25in]{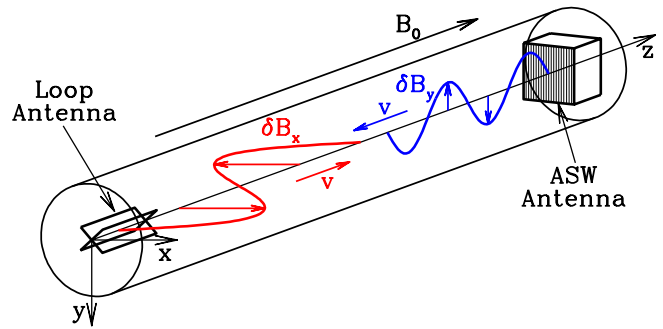}}
\caption{Schematic of the \Alfven wave turbulence experiment on the LAPD. 
The Loop antenna generates a large-amplitude \Alfven wave polarized in
the $x$-direction traveling up the mean magnetic field $B_0$ and the
ASW antenna generates a smaller amplitude \Alfven wave polarized in
the $y$-direction traveling down the mean magnetic field.
\label{fig:setup}}
\end{figure}

\begin{figure*}[bottom]
\hbox{
\hskip 1.2cm
 \hfill
\resizebox{3.0in}{!}{\includegraphics{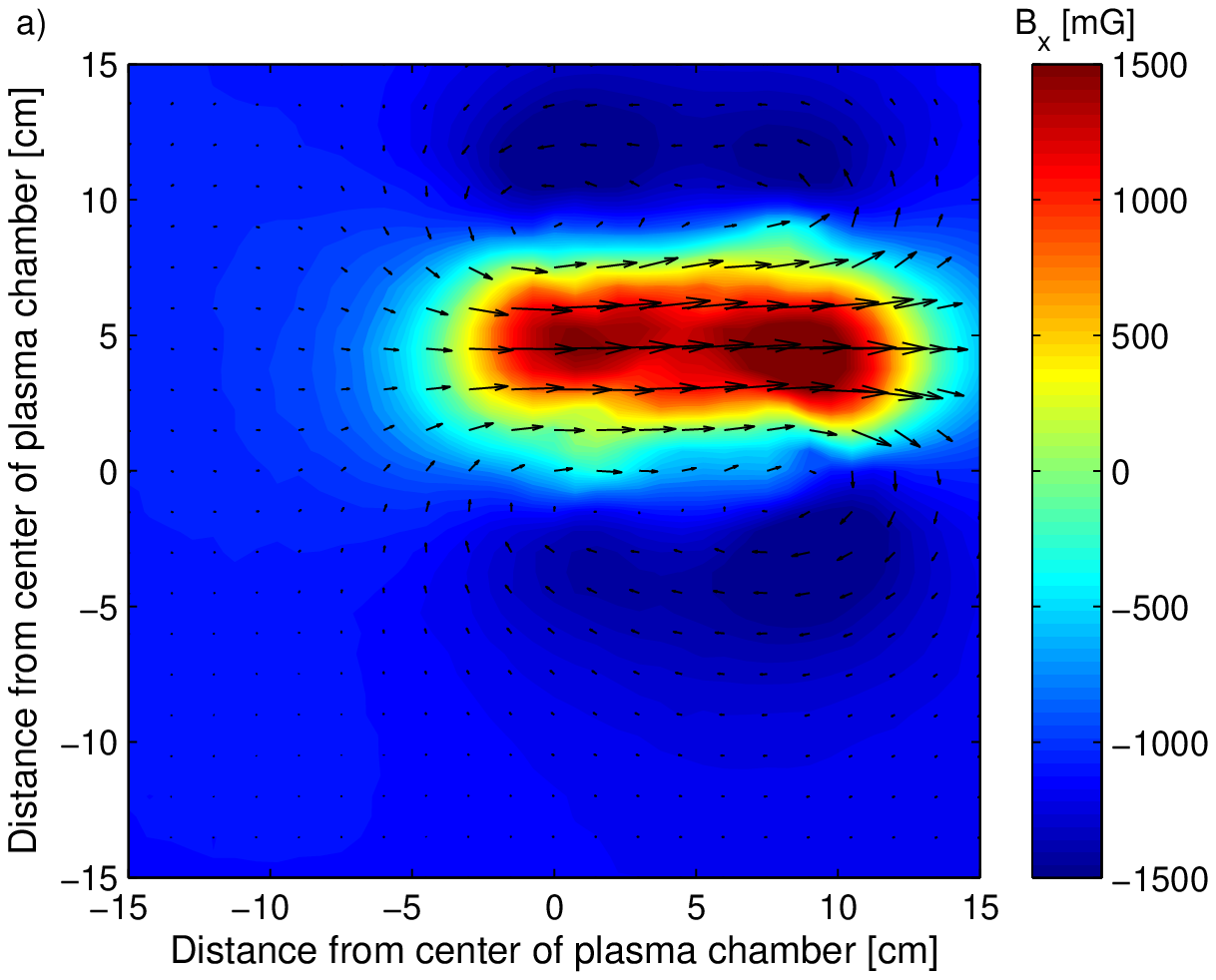}} \hfill
\resizebox{3.0in}{!}{\includegraphics{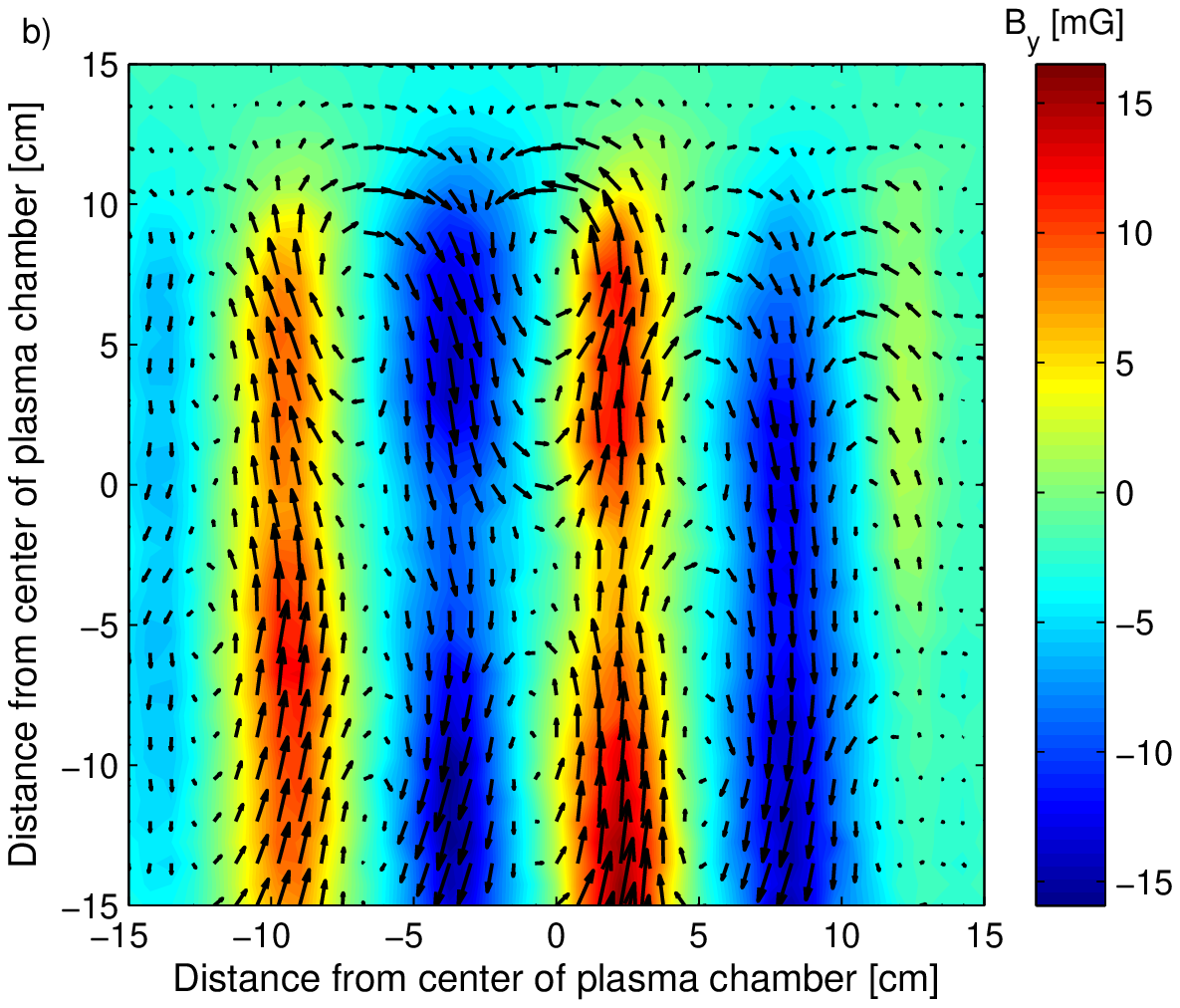}} \hfill}
\caption{Spatial magnetic-field waveforms  from the (a)  Loop antenna 
and (b) ASW antenna.  The colormap indicates (a) $\delta B_x$ for the
Loop antenna and (b) $\delta B_y$ for the ASW antenna in mG, and
arrows indicate the vector direction of $\delta \V{B}_\perp$.
\label{fig:antenna}}
\end{figure*}

The experimental setup is depicted in \figref{fig:setup}. At one end,
the Arbitrary Spatial Waveform (ASW) antenna
\citep{Thuecks:2009,Kletzing:2010} generates an \Alfven wave (blue)
with frequency $f=270$~kHz and a wave magnetic field polarized in the
$y$-direction, $\delta B_y$, characterized by a precise perpendicular
wavevector $\V{k}_{\perp 1}\rho_s= 0.16 \xhat$.  On the other end, a
Loop antenna \citep{Auerbach:2011}, constructed of two crossed current
loops phased to produce a dominantly horizontal wave magnetic field
$\delta B_x$, generates a large-amplitude \Alfven wave (red) with much
lower frequency $f=60$~kHz, characterized by a wavevector dominated by
a $y$-component $k_{\perp 2 y} \rho_s= 0.055 $.
The perpendicular wave magnetic fields, $\delta B_x$
and $\delta B_y$, are measured using an Els\"asser probe
\citep{Drake:2011} between the antennas approximately 2~m from the Loop
antenna and 9~m from the ASW antenna.  The measurements are taken over
a perpendicular plane of size $30$~cm by $30$~cm at a spacing of
$0.75$~cm, yielding a spatial grid of $41 \times 41$ measurement
positions. At each position, 10 shots are averaged to improve the
signal-to-noise ratio.

To generate a nonlinear interaction between the counterpropagating
Loop and ASW \Alfven waves, first the Loop antenna is turned on and
allowed to establish a steady wave pattern at $f=60$~kHz. Next, the
ASW antenna launches a perpendicularly polarized, counterpropagating
wave.  The spatial antenna waveform $\delta \V{B}_\perp$, measured in
single-antenna runs with identical timing, is presented in
\figref{fig:antenna}.  Colormaps show the magnitude of (a) $\delta
B_x$ for the Loop antenna and (b) $\delta B_y$ for the ASW antenna,
and arrows show the vector direction of $\delta
\V{B}_\perp$. The ASW antenna produces a waveform of 30~mG peak-to-peak amplitude 
with little $\delta B_x$ component, whereas the Loop antenna produces
a waveform of 3000~mG peak-to-peak amplitude that is dominated by the
$\delta B_x$ component.

\begin{figure*}
\hbox{
 \hfill
\resizebox{!}{2.4in}{\includegraphics*[0.35in,4.63in][7.4in,9.7in]{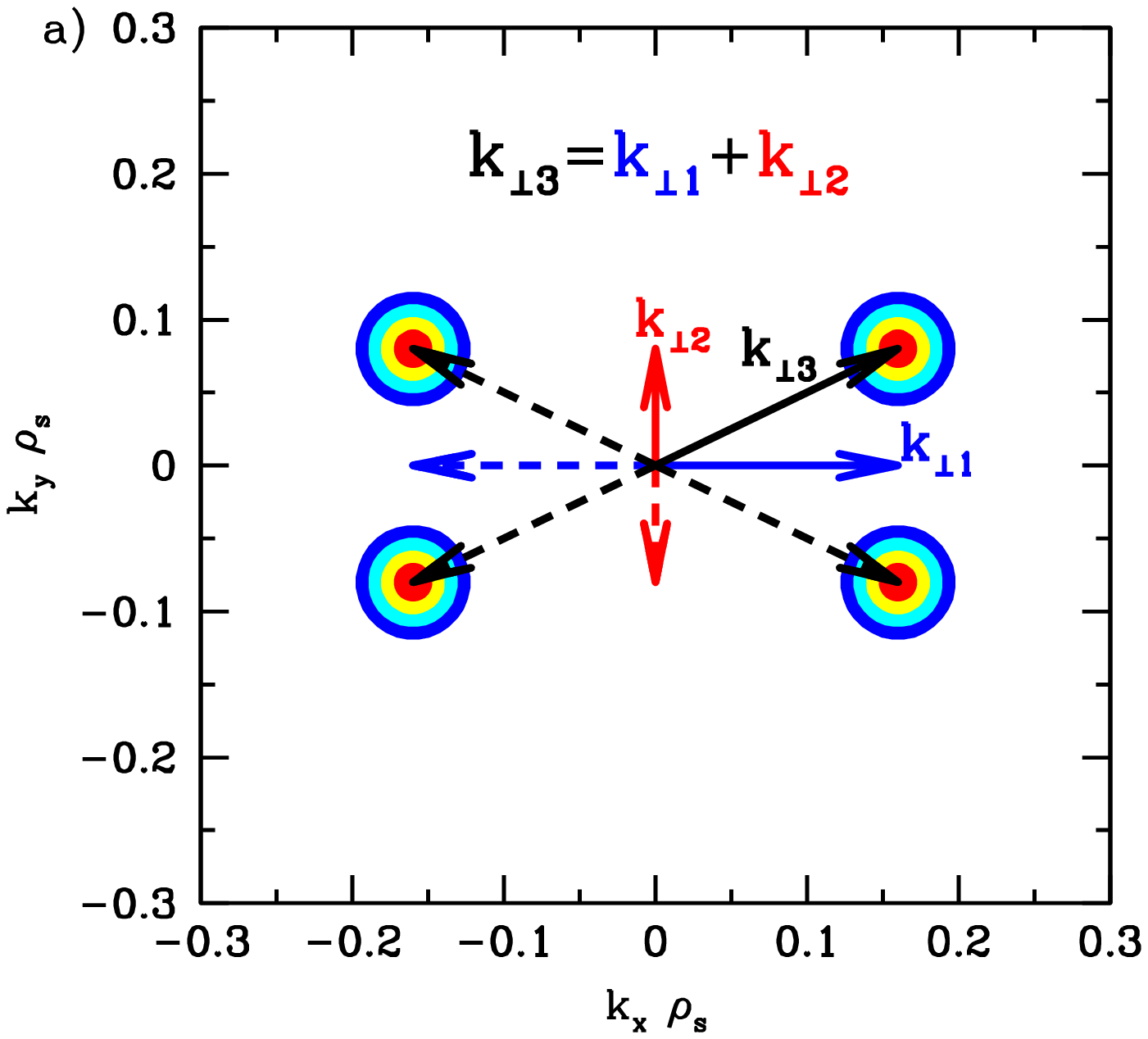}}
\hfill
\resizebox{!}{2.5in}{\includegraphics{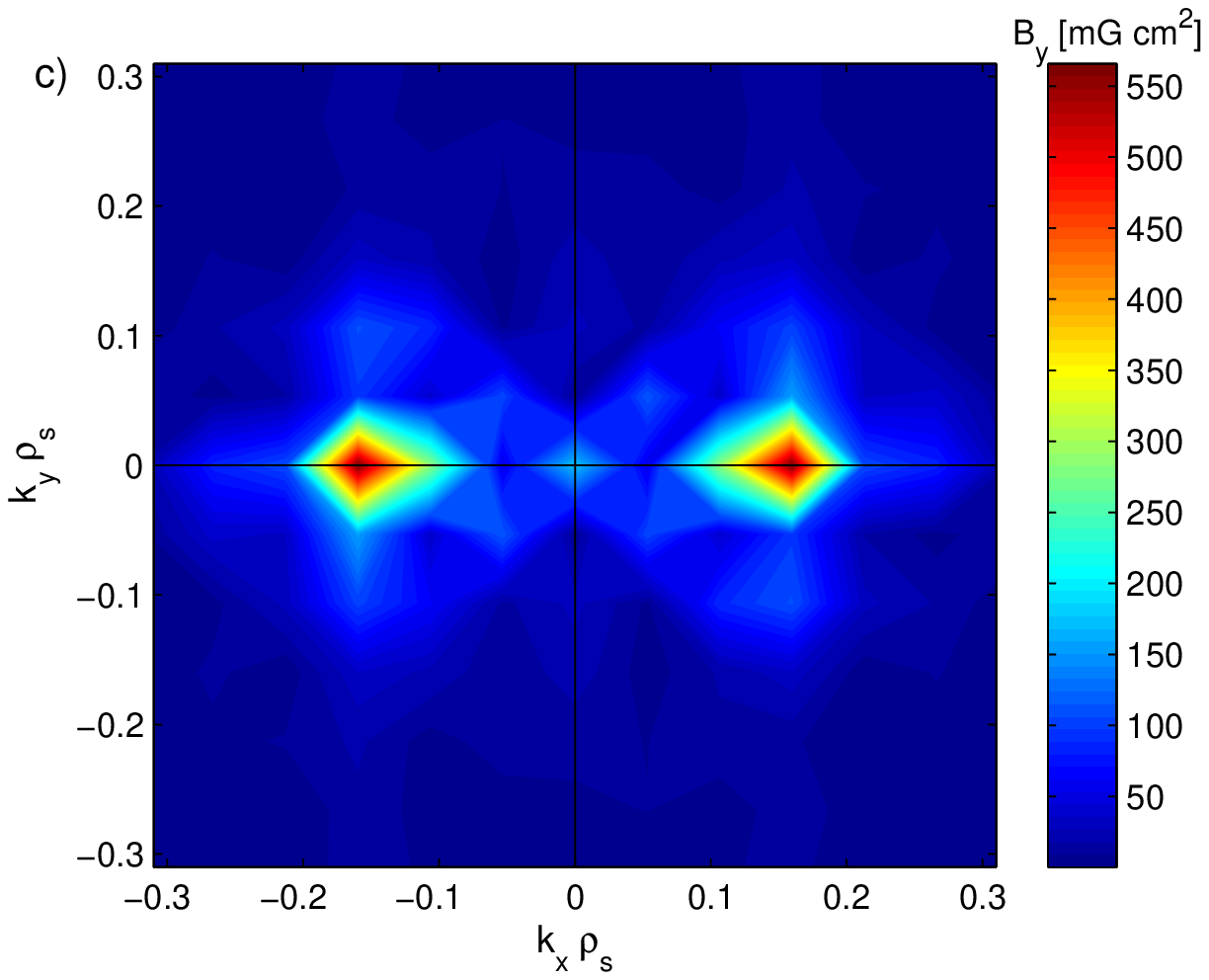}}
}
\hbox{
 \hfill
\resizebox{!}{2.5in}{\includegraphics{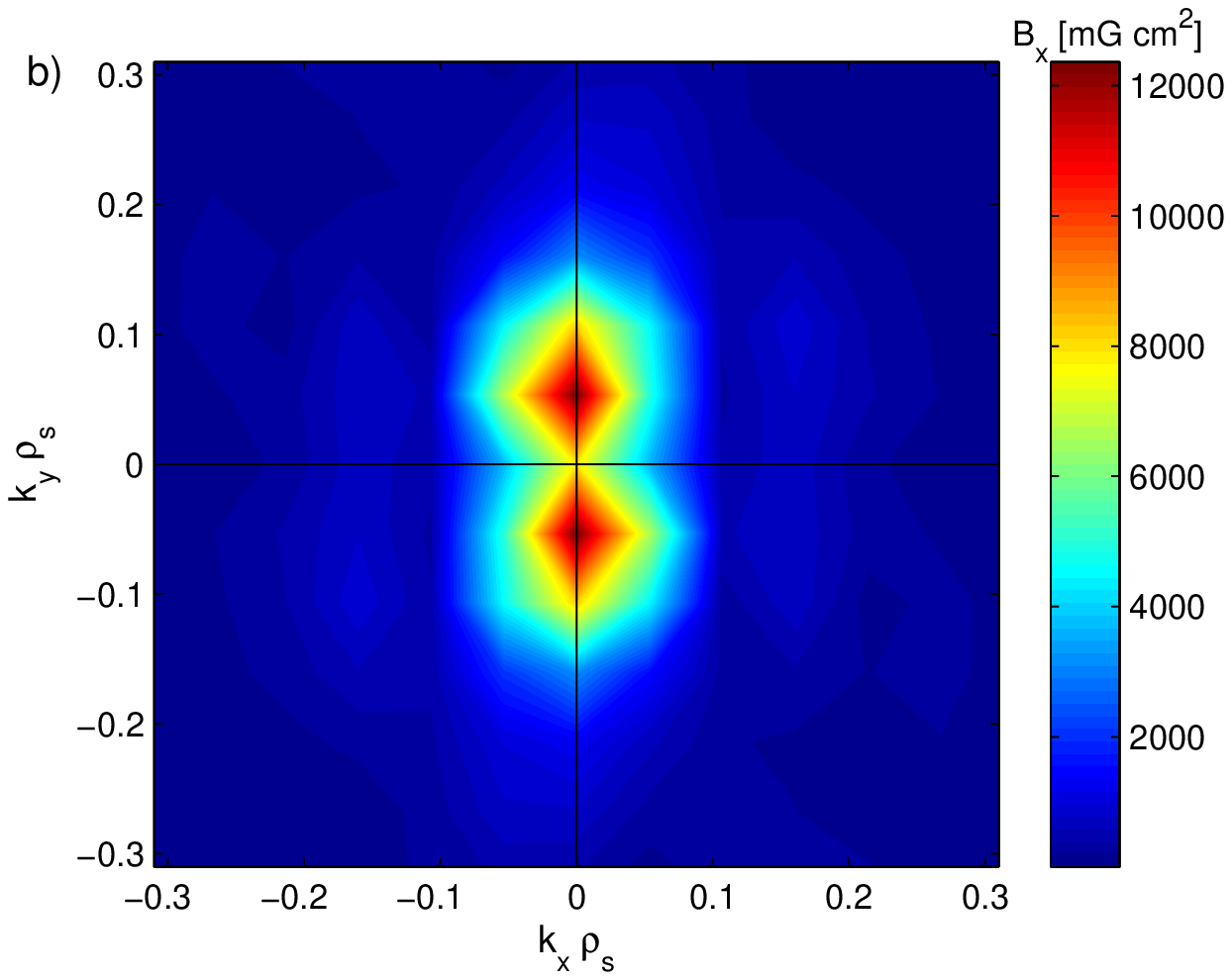}}
\resizebox{!}{2.5in}{\includegraphics{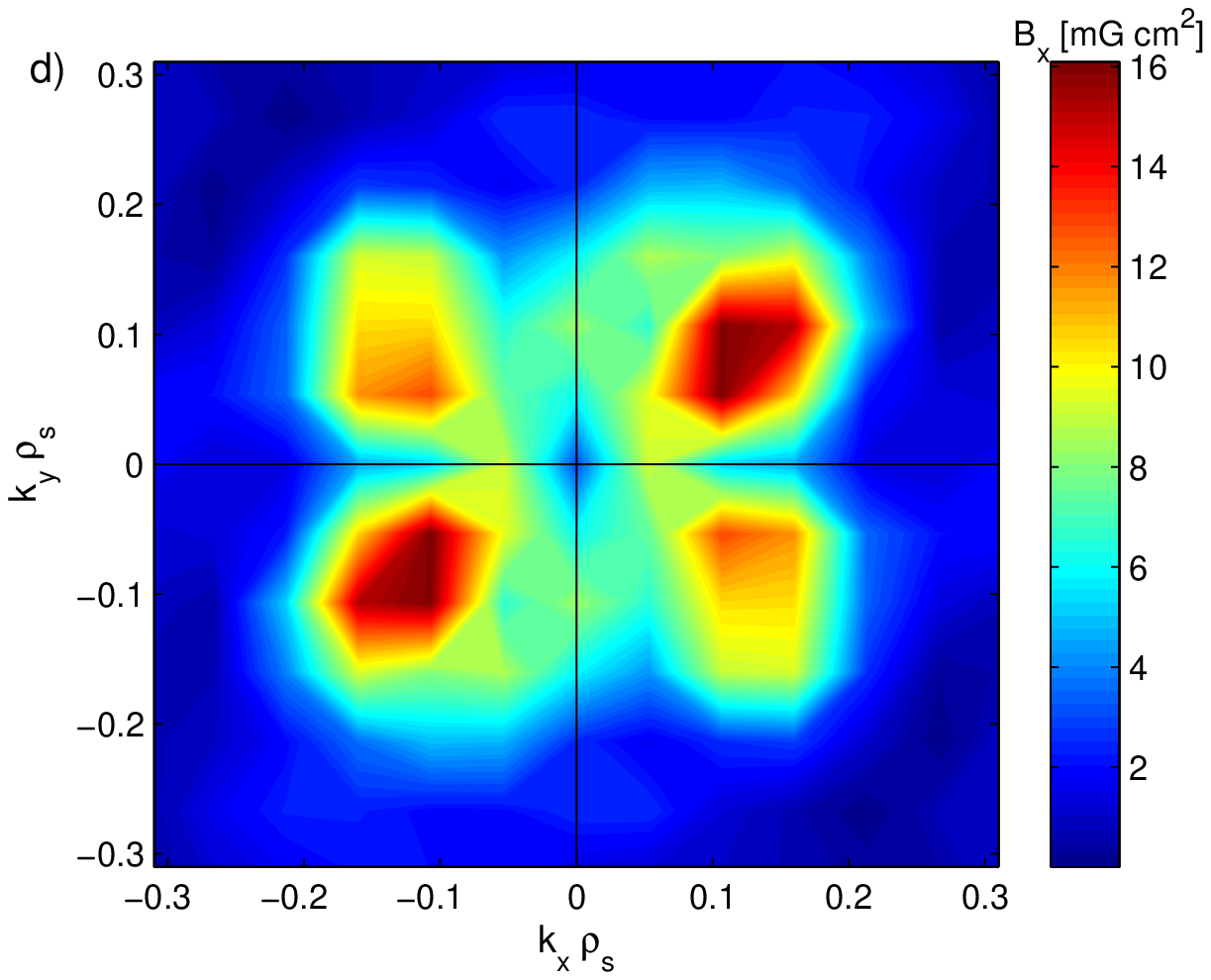}}
\hfill}
\caption{ (a) Diagram of $\V{k}_\perp$ for the ASW antenna (blue) 
and Loop antenna (red).  For the nonlinear daughter \Alfven wave,
$\V{k}_{\perp 3}$ (black) is the vector sum of the two antenna wave
vectors, $\V{k}_{\perp 3} = + \V{k}_{\perp 1} \pm \V{k}_{\perp 2} $
and $\V{k}_{\perp 3} = - \V{k}_{\perp 1} \pm \V{k}_{\perp 2}
$. Bullseyes indicate predicted power distribution of the nonlinear
product.  (b) Colormap of $\delta B_x(k_x,k_y)$ for the Loop
antenna by itself.  (c) Colormap (mG~cm$^2$) of $\delta B_y(k_x,k_y)$ for the ASW
antenna by itself. (d) Colormap of $\delta B_x(k_x,k_y)$ for the
nonlinear daughter \Alfven wave.
\label{fig:results}}
\end{figure*}

The experimental results are presented in \figref{fig:results}. We
plot (b) the Fourier transform in the perpendicular plane, $\delta
B_x(k_x,k_y,t)$, for the Loop antenna by itself and (c) $\delta
B_y(k_x,k_y,t)$ for the ASW antenna by itself.  Note that these are
the spatial Fourier transforms of the antenna patterns in
\figref{fig:antenna}, and  both antennas generate a pair of waves 
with $\pm \V{k}_{\perp}$.

The analysis of the counterpropagating \Alfven wave run is designed to
identify the nonlinear daughter \Alfven wave with two distinguishing
properties: (a) $k_{\parallel 3}= k_{\parallel 1}$, which means the
daughter wave will have the same frequency as the ASW wave, and (b) $
\V{k}_{\perp 3} =\V{k}_{\perp 1}+ \V{k}_{\perp 2}$.  First, to remove
the linear contribution to $\delta B_x$ caused by the large-amplitude
Loop \Alfven wave, we subtract the signal of $\delta B_x$ of the
Loop-antenna-only run from the corresponding signal of the
counterpropagating run to obtain $\delta B_x(x,y,t)$. Next, we Fourier
transform in time the interval with both waves to obtain $\delta
B_x(x,y,f)$. Since the signal of the daughter wave is expected to peak
at the same frequency $f=270$~kHz as the ASW wave, we bandpass filter
the $\delta B_x(x,y,f)$ signal over the frequency range 170~kHz~$\le f
\le$~370~kHz, setting the Fourier coefficients to zero outside this
range. The filtered frequency signal is then inverse Fourier
transformed from frequency back to time. Finally, we Fourier transform
in the perpendicular plane to obtain the spatial Fourier transform of
the daughter wave, $\delta B_x(k_x,k_y,t)$. 

The result of this procedure, shown in panel (d) of
\figref{fig:results}, is the key experimental result of this Letter.
The observational signature of the nonlinear daughter \Alfven wave is
clear, with the wave field $\delta B_x$ dominated by four wavevectors,
corresponding to all possible sums of perpendicular antenna
wavevectors, $\V{k}_{\perp 3}= +\V{k}_{\perp 1} \pm
\V{k}_{\perp 2}$ and  $\V{k}_{\perp 3}= - \V{k}_{\perp 1}  \pm
\V{k}_{\perp 2}$, as shown  schematically in panel (a)
of \figref{fig:results}. The noise level of the subtracted signal in
Fourier space is 1.2~mG cm$^{2}$, yielding a signal to noise ratio
S/N~$\gtrsim 10$, demonstrating that the measurement is physically
meaningful.  In addition, the amplitude of this nonlinear daughter
\Alfven wave, which agrees to order of magnitude with theoretical
expectations, peaks at the ASW wave frequency, $f_3=270$~kHz, as
predicted. These results demonstrate that we have successfully
measured the nonlinear interaction between counterpropagating \Alfven
waves, the fundamental building block of astrophysical plasma
turbulence.

This experimental finding verifies that the general properties of the
nonlinear interaction between counterpropagating \Alfven waves, as
derived theoretically in the idealized context of incompressible MHD,
hold even under the weakly collisional conditions relevant to many
space and astrophysical plasma environments. Although these conditions
formally require a kinetic description of the turbulent dynamics, our
results indicate that the key concepts derived from reduced fluid
models describe the essential nature of the turbulent interactions.
Future experiments will probe the turbulent dynamics at smaller
scales, providing guidance for the extension of existing turbulence
theories into the uncharted regime of kinetic turbulence at scales
below the ion sound Larmor radius, where the effects of wave
dispersion and kinetic wave-particle interactions influence the
turbulent dynamics.

Supported by NSF grants ATM 03-17310 and
PHY-10033446, DOE grant DE-FG02-06ER54890, NSF CAREER Award
AGS-1054061, and NASA grant NNX10AC91G.  The experiment presented here
was conducted at the Basic Plasma Science Facility, funded by the
U.S. Department of Energy and the National Science Foundation.

%
%

\end{document}